# A robust and lightweight deep attention multiple instance learning algorithm for predicting genetic alterations

Author: Bangwei Guo[1], Xingyu Li[2], Miaomiao Yang[3], Hong Zhang[2*], Xu Steven Xu[4*]

[1]Department of Applied Statistics, School of Data Science, University of Science and Technology of China;
[2]Department of Statistics and Finance, School of Management, University of Science and Technology of China;
[3] Clinical Pathology Center, The Fourth Affiliated Hospital of Anhui Medical University, Hefei, Anhui, China;
[4] Data Science/Translational Research, Genmab Inc., Princeton, New Jersey, USA;

## Abstract

Deep-learning models based on whole-slide digital pathology images (WSIs) become increasingly popular for predicting molecular biomarkers. Instance-based models has been the mainstream strategy for predicting genetic alterations using WSIs although bag-based models along with self-attention mechanism-based algorithms have been proposed for other digital pathology applications. In this paper, we proposed a novel Attention-based Multiple Instance Mutation Learning (AMIML) model for predicting gene mutations. AMIML was comprised of successive 1-D convolutional layers, a decoder, and a residual weight connection to facilitate further integration of a lightweight attention mechanism to detect the most predictive image patches. Using data for 24 clinically relevant genes from four cancer cohorts in The Cancer Genome Atlas (TCGA) studies (UCEC, BRCA, GBM and KIRC), we compared AMIML with one popular instance-based model and four recently published bag-based models (e.g., CHOWDER, HE2RNA, etc.). AMIML demonstrated excellent robustness, not only outperforming all the five baseline algorithms in the vast majority of the tested genes (17 out of 24), but also providing near-best-performance for the other seven genes. Conversely, the performance of the baseline published algorithms varied across different cancers/genes. In addition, compared to the published models for genetic alterations, AMIML provided a significant improvement for predicting a wide range of genes (e.g., KMT2C, TP53, and SETD2 for KIRC; ERBB2, BRCA1, and BRCA2 for BRCA; JAK1, POLE, and MTOR for UCEC) as well as produced outstanding predictive models for other clinically relevant gene mutations, which have not been reported in the current literature. Furthermore, with the flexible and interpretable attention-based MIL pooling mechanism, AMIML could further zero-in and detect predictive image patches.

# Introduction

The detection of genetic alterations aims to achieve precision treatment of cancer[1], including conventional chemotherapy and targeted therapies[2,3]. However, molecular and genetic tests can be time consuming, expensive, and difficult to implement at scale[1]. Automatic computer-aided diagnostics can reduce pathologists' workloads and diagnosis mistakes[4]. Compared with genomics data, H&E-stained WSIs are ubiquitously available with the rise in whole slide scanner technology[5], and can reflect morphological changes in tumor cells and their microenvironment in extremely detailed, gigapixel resolution[6]. Many studies[5,7-9] have revealed that deep learning can learn unique morphological features just in WSIs to robustly infer the genotype and elucidate biological mechanisms of downstream effects of molecular alterations across many cancer types, such as lung cancer[2], bladder cancer[10], and colorectal cancer[11-13].

WSIs are large with Giga pixels[2], so it is extremely challenging to process on a General Processing Unit (GPU) at once. Currently, WSIs are usually split into small image patches to train neural networks to predict gene mutations[1,2,11,12,14]. As a result, many image patches are present for each WSI, while only one mutation status is available for that patient (or slide). Therefore, features from multiple patches in a WSI must be aggregated to predict the mutation status for that WSI.

The published aggregation strategies in the field of digital pathology are commonly done with the idea of Multiple Instance Learning (MIL). Each WSI is considered as a bag with multiple instances (image patches). The aggregation methods can be roughly divided into two general categories[15-17]: instance-base and bag-based methods, namely aggregating patch predictions and aggregating patch feature, respectively. So far, aggregating patch predictions has been the mainstream strategy for predicting genetic alterations using WSIs in the current literature[1,2,12-14]. In patch prediction aggregation, patch-level predictions are first obtained either via a fully trained deep Convolutional Neural Network (CNN) classifier from scratch[1,2] or transfer learning using pretrained models[14,18], they are then aggregated into patient-level prediction using different statistics, including the average aggregation[2], a majority vote[1,13], quantile aggregation[19], or selecting the top-ranked patches by iteration[12].

Aggregating patch features is an alternative strategy, which has been widely used in classification models in the field of digital pathology. This method usually achieved a better performance for tasks where global (i.e., bag-level) predictions are more important[20]. Consequently, it is very suitable for analyzing WSIs when researchers can only obtain slide-level labels[20]. Coupling with deep learning, aggregating patch features have been used to identify informative patches of WSIs. Successive 1-D convolutional layers are commonly used to aggregate the features of a tile to derive a tile score[21-24]. It is well accepted that maximum scoring tiles often carry the most predictive information. Therefore, large number of the studies[21,22,24,25] adapted max-pooling (maximum scoring patches) to identify predictive patches and further aggregate the tile level information for classification. Recently, Durand et al.[23] proposed WELDON, which selected and aggregated the highest and lowest scoring patches for classification, and demonstrated an improved performance when compared to max-pooling and mean-pooling. Furthermore, Courtiol et al. modified the WELDON model and proposed

CHOWDER to classify breast cancer based on whole slide images[22]. Schmauch et al.[26] proposed a simple variant of the max-pooling model called HE2RNA, which randomly sampled top-k scoring patches from a list of prespecified values to improve model generalization and prediction of RNA-seq expression. So far, despite of wide application of feature aggregation strategy in digital pathology in general, relatively limited work has been done using this strategy to predict genetic alterations.

Another popular approach for aggregating patch features is the self-attention mechanism, which calculates an attention weight for each patch and assign higher weights to more informative patches. The patch features are then aggregated as a weighted sum to obtain slide-level prediction (Ilse et al.[27]; Lu et al.[28]; Yao et al.[4]; Saillard et al.[24]). However, so far, relatively few attention-based algorithms have been developed to predict gene mutations. Anand et al.[29] used an attention-based DNN (deep neural network) model to predict BRAF mutation in thyroid cancer. Qu et al.[30] extracted the features of tumor patches in breast carcinoma using ResNet101 and applied attention-based aggregation to predict genetic mutations and biological pathways. The main challenges for attention-based algorithms have been the model complexity and their data-hungry nature[13,25].

In this paper, we propose a novel, attention-based feature aggregation algorithm to predict gene mutations from WSIs. Our model was consisted of a decoder and a residual weight connection, which can restore the 1-D scores to features of a suitable dimension to reduce the potential loss of information caused by the 1-D convolutional layers used in the current existing algorithms and allow for incorporating attention mechanism that is impossible for a one-dimensional score. Unlike other attention-based models, we only used a small number of representative patches through an encoder architecture. This can greatly reduce the complexity of attention-based models and consequently circumvent the challenges like long operation time and large GPU memory usage generally related to attention mechanism, and allow for easy implementation of attention mechanism for prediction models for genetic mutations. With the advantages of feature aggregation using 1-D scores and a lightweight attention mechanism, our model can further zero in on the most important patches of WSIs, and improve model interpretability.

## Method

The workflow of our proposed algorithm is illustrated in **Figure 1**. **Figure 1a and 1b** show the data preprocessing, including feature extraction pipeline and optimal clustering, while **Figure 1c** illustrates the structure of our proposed algorithm, AMIML.

### Dataset

Anonymized scanned WSIs of diagnostic tissue slides (FFPE tissue) stained with hematoxylin and eosin were downloaded in SVS format from The Cancer Genome Atlas (TCGA) project through the Genomic Data Commons Portal (https://portal.gdc.cancer.gov/) with matched genomic data. The WSIs for four solid tumor types: uterine corpus endometrial carcinoma (UCEC), breast invasive carcinoma (BRCA), glioblastoma multiforme (GBM), and kidney renal clear cell carcinoma (KIRC) were analyzed in this study.

**Feature extraction**

For each WSI, we first generated its thumbnail, then apply the OTSU[31] algorithm, which could automatically generate the best segmentation threshold based on the image. Then we applied the U-Net[32] structure to segment tissues of WSIs and exclude background areas. After that, WSIs were tessellated into non-overlapping patches with a size of 224 × 224 pixels. These patches were then color normalized using Macenko's method[33], which could improve classifier performance for mutation prediction[1].

We applied a fine-tuned Xception[34] model which was a supervised trained CRC (Colorectal Cancer) tissues[35] classifier and was proved very effective in our previous work[36,37] to extract features. Each tile input to the network was extracted as a 256-dimensional feature. The pipeline for feature extraction is illustrated in **Figure 1a**.

**Pre-selection of patches**

Tumor tissues are usually selected for development of prediction algorithms for genetic alterations (Kather et al.[1] and Coudray et al.[2]) since it is commonly assumed that tumor regions carry the most predictive information. Recently, we proposed an unsupervised clustering method to optimize the prediction of gene mutations, which could provide a better predictive performance compared to approaches solely based on tumor regions on the WSI slides[38]. Briefly, for each of the four cancer types, we used K-means method to group all the patches into four clusters according to the different morphologic features. Only WSIs which have more than 10 patches on each cluster were included in this study. In total, 1946 WSIs (UCEC:388, BRCA:728, GBM:507, KIRC:323) from 1647 patients (UCEC:388, BRCA:728, GBM: 214, KIRC:317) were included in this study. The cluster with the best predictive performance was selected for a particular gene. The workflow for patch selection using unsupervised clustering is illustrated in **Figure 1b.**

**MIL models description**

After features were extracted, we first applied the idea of MIL and packed the features of each patient's WSIs into a "bag"-$n$ ×256 feature matrix, and the 256-D feature of each tile was regarded as an instance. Traditional MIL aggregation defined a positive bag as having at least one positive instance and a negative bag as one with all negative instances[16]. However, WSI contains a large number of patches, and it is obviously unreasonable to use this traditional assumption, which is also demonstrated in our experiments. In general, some representative patches were taken in the bag and their output were aggregated as the final bag output[21,22,24,27,36,39].

We developed Attention-based Multiple Instance Mutation Learning (AMIML), for which the detailed framework is shown in **Figure 1c.** AMIML is comprised of four important components: (1) an encoder, (2) a decoder, (3) a residual weight connection, and (4) Attention mechanism as follows:

(1) The encoder is successive 1-D convolutional layers, through which the 256-dimensional features are sequentially reduced to 1-D scores. Then, according to the scores of the patches, the model selected the patches with the top R highest and lowest scores as CHOWDER does, and R is a hyperparameter. The encoder is illustrated below:

$$\chi_{n_{tiles}\times 256} \xrightarrow{\text{1D convolutional layers}} E(\chi)_{n_{tiles}\times 8} \xrightarrow{\text{score}} E(\chi)_{n_{tiles}\times 1} \xrightarrow{\text{select}} E(\chi)_{2R\times 1}$$

(2) The decoder is aiming to decode the features reduced to 1-D scores by the encoder into features of higher-dimensional features. We can explain it with a simple formula: $E(\chi)_{2R\times 1} \xrightarrow{\text{Decoder}} D(\chi)_{2R\times 8}$. The score part gives not only the patches with the highest and lowest scores, but also the position indexes of these patches in the bag. This allows us to find a corresponding subset of $E(\chi)_{n_{tiles}\times 8}$: $E(\chi)_{2R\times 8}$. In theory, perfect encoder and decoder will make sure that $E(\chi)_{2R\times 8} = D(\chi)_{2R\times 8}$, but it is not practically possible.

(3) In order to make the result decoded by the decoder as close to $E(\chi)_{2R\times 8}$ as possible, we use the idea of residual layer[40] and introduce residual weight connection:

$$E(\chi)_{n_{tiles}\times 8} \xrightarrow{\text{Encoder}} E(\chi)_{2R\times 1} \xrightarrow{\text{Decoder}} D(\chi)_{2R\times 8} \xrightarrow{\text{Residual weight connnection}} \alpha D(\chi)_{2R\times 8} \oplus E(\chi)_{2R\times 8}$$

where $\alpha$ is a hyperparameter weight that aims to reduce the impact of $D(\chi)_{2R\times 8}$ on the overall result to make $\alpha D(\chi)_{2R\times 8} \oplus E(\chi)_{2R\times 8}$ and $E(\chi)_{2R\times 8}$ as close as possible.

(4) We use the self-attention mechanism consisting of three key weight matrices: query matrix Q, key matrix K and value matrix V[41]. We design three FC layers respectively to obtain the weight matrices and calculate the contribution weights of patches, and assign higher attention weights to important patches in the bag:

$$Attention(Q, K, V) = softmax(\frac{QK^T}{\sqrt{d_k}})V$$

Finally, we average the weighted two-dimensional vector of each instance and then make probability normalization with the SoftMax function to predict the mutation probability of the patient.

**Comparison with Baseline Algorithms**

For comparison, we conducted experiments on the same data using five different methods: CHOWDER[22], HE2RNA[26], MIL mean-pool[27], MIL max-pool[27], and patches mean-pool aggregation method[14]. And these detail frameworks are shown in **0. 错误!未找到引用源。**

**Patient-level cross-validation**

We randomly split the patient cohort into 5 partitions (training set: validation set: test set =3:1:1) for each gene, and both mutant and wild-type individuals follow this ratio to keep the target labels balanced between partition. The models for predicting mutations were trained on the training set, and the model with the lowest loss on the validation set was selected, subsequently evaluated on the test set. For tile-level training methods, we make sure that all patches inherited the mutation label of the corresponding patient. In addition, all the patches from a patient were assigned to either the training, validation, or test set. To address the problem of imbalanced number of classes, we used a weighted loss function.

**Neural network training and hyperparameter optimization**

We fixed R=100 and $\alpha$=0.1 in the AMIML model and the CHOWDER model. The setting of hyperparameters during the training process was as follows. The initial learning rate is 0.01, and the learning rate was multiplied by 0.5 if the loss on the validation set did not decrease after 10 consecutive epochs; The batch size was 64; The initial epoch number was set to 300 with the early-stopping strategy; Cross entropy loss function and Adam optimizer were used. We implemented the networks in PyTorch and trained them on an NVIDIA GeForce GTX 3090 Ti GPU.

# Results

**Comparison with five baseline algorithms**

*UCEC*

**Figure 2** illustrates the average AUC values from the 5-fold cross validation for the AMIML algorithm and five baseline prediction models based on image patches from the TCGA UCEC cohort. AMIML apparently outperformed the five baseline algorithms and provided the best predictive performance for five out of the six clinically relevant genes for UCEC patients (i.e., MTOR, POLE, ATM, JAK1, and TP53). For PTEN mutation, although MIL-mean and Patch-mean approaches had the highest AUC (0.78), AMIML delivered decent predictive performance with an AUC value of 0.76. Generally, HE2RNA and MIL-max had inferior performance compared to all other algorithms, while the predictive performance of CHOWDER, MIL-mean, and Patch-mean varied from gene to gene.

*BRCA*

**Figure 3** shows the average AUC values from the 5-fold cross validation for the TCGA BRCA cohort. Similarly, AMIML had the best predictive performance in the majority of the tested genes (four out of six) compared to the five baseline algorithms (i.e., BRCA1, BRCA2, ERBB2, and PIK3CA). For the other two genes (CDH1 and TP53), the performance of AMIML was close to best one (i.e., CDH1: 0.74 for AMIML vs. 0.75 for CHOWDER; TP53: 0.77 for

AMIML vs. 0.78 for MIL-mean and Patch-mean). For the PTEN mutation, although MIL-mean and Patch-mean approaches had the highest AUC (0.78), AMIML delivered decent predictive performance with an AUC value of 0.76. Although CHOWDER demonstrated the best predictive performance in CDH1, its performance varied for different genes and was the worst in predicting BRCA2 mutation. Similarly, HE2RNA remained the worst in comparison with the other tested algorithms, while the predictive performance of MIL-max, CHOWDER, MIL-mean, and Patch-mean varied for individual genes.

*GBM*

Figure 4 shows the average AUC values from the 5-fold cross validation for the TCGA GBM cohort. Probably due to the relatively small sample size of the GBM cohort, the predictive performance of all the tested algorithms appeared less stable compared to the other studied cancer types. Nevertheless, AMIML still demonstrated the best predictive performance in three of the six tested genes (AUCs: 0.72 for RB1, 0.83 for IDH1, and 0.78 for KMT2C). Once again, HE2RNA and MIL-max tended to produce suboptimal predictive performance. For example, HE2RNA had only an AUC of 0.38 for RB1 mutations in comparison of 0.72 for AMIML. Both MIL-mean and Patch-mean algorithms provided the worst performance in predicting ZFHX3 mutation in GBM. CHOWDER best predicted ZFHX3 (AUC = 0.73), KMT2C (AUC = 0.78), and TRRAP (AUC = 0.78).

*KIRC*

Figure 5 shows the average AUC values from the 5-fold cross validation for the TCGA KIRC cohort. AMIML remained the most robust algorithm in KIRC, and provided the best predictive performance in five (i.e., ATM, SETD2, BAP1, KMT2C, and TP53) out of the six selected genes among the studied algorithms. For PBRM1, the AUC based on AMIML was only 0.02 points lower than the best algorithm (AUC = 0.65 for MIL-mean). HE2RNA and MIL-max remained the worst prediction algorithms for gene mutations for KIRC. For SETD2 mutation, both HE2RNA and MIL-max could not identify any association between H&E image features and the mutation probability (AUC = ~ 0.5), while using AMIML, the prediction AUC was 0.7, suggesting a strong correlation. Similar situation happened for KMT2C as well, where HE2RNA only produced an AUC of 0.56, whereas AMIML delivered an outstanding AUC of 0.73. The predictive performance of CHOWDER was generally comparable to or slightly lower than that for MIL-mean and Patch-mean in KIRC.

**Comparison with the published models**

For the studied genes, we compared the predictive performance of AMIML with the published results based on TCGA cohorts. **Table 1** illustrates the average AUC values based on AMIML and reported by current literature. For the BRCA dataset, current publications did not find any clinically useable predictive values of H&E images for both BRCA1 and BRCA2 genes as the reported AUC was merely above 0.5[1,30]. AMIML drastically improve the prediction of BRCA1 and BRCA2 from the vicinity of 0.5 to 0.73 and 0.65, respectively. In addition, AMIML significantly improved the prediction for ERBB2 mutation, i.e., the AUC based on AMIML was 0.72, compared to approximately 0.63 reported in the current literature[1,30]. The prediction of PIK3CA mutation in BRCA cohort was also slightly better for AMIML than reported data[1,30].

For the TCGA-KIRC dataset, Kather et al. reported the AUC values for five out of the six selected clinically relevant gene mutations based on a deep-learning model (**Table 1**)[1]. AMIML provided marked improvement of prediction for three genes (KMT2C, TP53, and SETD2). For KMT2C and TP53, the AUC was improved from approximately 0.6 to 0.7, representing an improvement of ~17%. No literature data are currently available for BAP1 mutation. AMIML provided an admirable prediction of BAP1 mutation with an AUC of 0.7. The AUC values for ATM and PBRM1 were similar between AMIML and reported deep-learning models.

With regards to UCEC, Hong et al.[42] reported deep-learning models for gene mutations based on combined TCGA and CPTAC datasets. Our AMIML model had apparently a higher AUC value (0.71) for predicting JAK1 mutation than that from Hong's model (0.61). In addition, AMIML models for POLE and MTOR provided a better prediction than reported models. The predictions based on AMIML for PTEN and ATM mutations were similar between AMIML and reported models, whereas Hong's model for TP53 appeared to have higher AUC (0.87) than AMIML (0.76).

Few studies have been conducted for gene mutations in GBM. Cui et al[43] developed a model for IDH1 mutation using both GBM and LGG datasets. Both Cui's model and our AMIML provide an outstanding prediction for IDH1 mutation with an AUC around 0.83. In addition, our study reveals that AMIML-based deep-learning models could provide excellent predictions (AUC > 0.72) for several clinically important gene mutations in GBM (i.e., TRRAP, KMT2C, ATEX, RB1, and ZFHX1). No deep-learning models based on H&E images for prediction of mutation of these genes have been developed.

**Attention-detected relevant patches**

Attention mechanism is designed to assign higher weights to the patches that are more relevant to the outcome. This feature allows us to identify predictive image patches. For illustration purpose, we exhibit two examples for IDH1 gene in GBM (错误!未找到引用源。) and BAP1 gene in KIRC (错误!未找到引用源。**7**).

IDH1 gene is an important tumor molecular marker in glioblastoma[44]. For patients with IDH1 mutations, AMIML detected morphologic features like cellular abundance, nuclear pleomorphism, and active mitotic activity, as well as identified important patches showing complex and disordered proliferative vascular components or coagulation-type tumor necrosis (**Figure 6**), consistent with the morphological features reported for IDH1-mutant tumor[45]. In addition, IDH1-mutant glioblastoma can develop from IDH1-mutant anaplastic astrocytoma. The patches of high attention weights also exhibited similar histology. For IDH1 wild-type patients, the selected image patches showed smaller cancer cells.

For kidney cancer (renal cell carcinoma in particular) patients, BAP1 gene mutations are associated with high-grade tumors and poor prognosis[46], the tumor tissues with mutations of the BAP1 gene (e.g., loss of BAP1) were more likely to present with high-grade tumors and showed specific features: abundant acidophilic cytoplasm, eccentric nuclei and prominent macronucleoli[46](**0**), whereas lower-grade cancer cells often exhibit more transparent cytoplasm and smaller nucleoli[47]. Not surprisingly, these features have been captured by AMIML in the patches of high attention weights. BAP1 mutants presented features similar to those for high-grade tumors while BAP1 wild types had characteristics resembling the low-grade cancer.

## Discussion

Gene mutation detection by digital WSIs is an important problem in computational pathology. In recent years, many studies have demonstrated that deep learning can extract subtle visual features from histological images, which can be used to predict molecular alterations. The existing deep-learning algorithms are mainly instance-base methods (i.e., aggregating patch predictions) despite that bag-based models (aggregating patch feature) using 1-D convolutional layers (e.g., WELDON, CHOWDER, and HE2RNA, etc.) have been proposed for other digital pathology applications. In addition, although self-attention mechanism-based models become increasingly popular, their complexity and data-hungry nature pose a significant challenge on its utility in predicting genetic alterations[13].

In this paper, we proposed a lightweight attention-based MIL model (AMIML) to predict gene mutations in H&E-stained whole-slide images. Unlike the traditional bag-based models (e.g., WELDON, CHOWDER, and HE2RNA, etc.) that utilize successive 1-D convolutional layers, our model integrated the 1-D convolutional encoder with a decoder and a residual weight connection, so that the encoded 1-D scores can be restored back to features to mitigate the potential loss of information during 1-D convolutional encoding process for the traditional bag-based models. In addition, the restored features with reduced dimension allowed for further integration of a lightweight attention mechanism in the model following the 1-D encoding, and thereby not only reducing the computational complexity for attention-based models but also facilitating further identification of relevant patches related to target mutations. We compared our proposed AMIML with five published popular algorithms using mutation data for 24 clinically relevant genes from four TCGA cancer datasets (i.e., UCEC, BRCA, GBM and KIRC). In addition, the prediction performance of AMIML (AUC values) was compared with that reported in the recent literature.

The present study demonstrated the robustness of our proposed AMIML algorithm. AMIML provided the best prediction for the vast majority of the tested genes (17 out of 24 studied genes). For the other 7 genes, even though AMIML did not provide the best prediction, its performance was usually still in the neighborhood of the best approaches, demonstrating excellent robustness. On the other hand, our study revealed that max pooling algorithms such as MIL-max and HE2RNA generally produced inferior performance, and therefore are not recommended for predicting gene mutations. Although CHOWDER, mean-based approaches like MIL-mean and Patch mean algorithms often produce acceptable predictions, they still occasionally failed to provide a satisfactory prediction performance (e.g., the lowest AUC values among the tested algorithms).

Compared to the existing mutation models for the genes from the four cancer types, our AMIML algorithm resulted in significant improvement for predicting a wide range of genes in these cancers. For the KIRC dataset, AMIML markedly improved prediction of three genes (KMT2C, TP53, and SETD2). In kidney cancer, TP53 is the most frequently mutated tumor suppressor gene (32%), while SETD2 gene is located on the same short arm of chromosome 3 where the VHL gene is also located, and is associated with more aggressive disease and poor prognosis[46]. In breast cancer, BRCA1 and BRCA2 mutations are highly predictive of treatment effect of PARP inhibitors such as olaparib and rucaparib, the response rate tended to be higher in patients with BRCA1/BRCA2 mutations, and testing of BRCA1/BRCA2 is NCCN recommended[48]. So far, none of the existing models have been able to predict mutations for BRCA1 and BRCA2 (AUC ~ 0.5; [1]). AMIML drastically improved the prediction of BRCA1 and BRCA2 mutations from an AUC of 0.5 to 0.73 and 0.65, respectively. In addition, AMIML significantly improved the prediction for ERBB2 mutation from an AUC of 0.63 in the current literature[1,30] to 0.72. In endometrial cancer, AMIML outperformed existing models for prediction of JAK1, POLE, and MTOR mutations[42]. Furthermore, we also developed AMIML-based deep-learning models using H&E images for the prediction of mutations that have not been reported in the current literature including BAP1 in kidney cancer, and TRRAP, KMT2C, ATEX, RB1, and ZFHX1 for GBM. AUC of 0.7 or higher was achieved for all these genes. AMIML also showed a good predictive performance on the TP53 gene in pan-cancers, which is an important gene with a high mutation rate in various cancers and closely related to prognosis[18,49]. The predicted AUCs of the TP53 gene mutation in UCEC, BRCA, and KIRC are 0.758, 0.767, and 0.731, respectively.

The AMIML workflow proposed in this work is consisted of three mechanisms to identify predictive image patches. First, an unsupervised clustering was utilized in the pre-processing step to detect image patches that can better predict the mutation of individual genes. In addition, similar to WELDON and CHOWDER, AMIML selected the patches with the top R highest and lowest scores for a prediction model development. These image patches with the highest and lowest scores are believed to carry the most predictive information. Moreover, with the flexible and interpretable attention-based MIL pooling mechanism, AMIML can further zero in on the patches and ROIs that are more relevant for predicting mutations for further analysis by pathologists.

# Acknowledgements

The research of Bangwei Guo, Xingyu Li, and Hong Zhang was partially supported by National Natural Science Foundation of China (No. 12171451) and Anhui Center for Applied Mathematics.

# Author contributions

X.S.X., B.G., X.L., and H.Z. contributed to design of the research; X.L., B.G., and X.S.X. contributed to data acquisition; B.G., X.L., and X.S.X. contributed to data analysis. B.G., X.L., X.S.X., M.Y., and H.Z. contributed to data interpretation. B.G., X.S.X., H.Z., M.Y. and X.L. wrote the manuscript; and all authors critically reviewed the manuscript and approved the final version.

# Data availability

The TCGA dataset is publicly available at the TCGA portal (https://portal.gdc.cancer.gov).The public TCGA clinical data is available at the website(https://xenabrowser.net/datapages/) and the website(https://portal.gdc.cancer.gov). Xception model weights are available at (https://github.com/fchollet/deep-learningmodels/releases/download/v0.4/xception_weights_tf_dim_ordering_tf_kernels_notop.h5)

# Code availability

Source code is available at https://github.com/Boomwwe/AMIML.

**Figure 1. (a) The workflow of data preprocessing** (I) Images of four cancers were first downloaded from TCGA in the SVS format; (II) Each whole-slide H&E image was preprocessed to remove the background areas using a U-net, split into non-overlapping tiles with a size of 224 x 224 pixels, and color normalized; (III) A fine-tuned Xception model-based feature extractor was used to generate patch features; (IV) The features of each patient were sampled into a bag. **(b) The optimal clustering method proposed in our previous work[38].** Features of all patches were clustered into four clusters by the k-means method, and the cluster with the best AUC was selected. **(c) The architecture of AMIML for predicting gene mutations.** AMIML is comprised of four important components: (1) an encoder, (2) a decoder, (3) a residual weight connection, and (4) a self-attention mechanism. Through the encoder structure similar to Chowder, our model selected the R patches with the highest and lowest scores for predicting mutation status. Then, through the decoder and residual weight connection structure, these scores were reduced to corresponding higher-dimensional representative features, and through the self-attention module our model calculated the contribution weight of patches to the bag output and finally aggregated them into the patient-level prediction. This figure describes the case of R=2. More detailed description is shown in the Method section.

**Figure 2. Comparison of the proposed AMIML algorithm with five baseline prediction models based on image patches from the TCGA UCEC cohort.** This bar plot shows the AUC values of the 5-fold cross-validations for six models. The lengths of the color bars shown on the right-hand sides represent the average AUC values of 5-fold cross-validations, and the error bars represent the standard deviations of the AUC values of the 5-fold cross-validations.

**Figure 3. Comparison of the proposed AMIML algorithm with five baseline prediction models based on image patches from the TCGA BRCA cohort.** Refer to the caption of Figure 2 for detailed descriptions.

**Figure 4. Comparison of the proposed AMIML algorithm with five baseline prediction models based on image patches from the TCGA GBM cohort.** Refer to the caption of Figure 2 for detailed descriptions.

**Figure 5. Comparison of the proposed AMIML algorithm with five baseline prediction models based on image patches from the TCGA KIRC cohort.** Refer to the caption of Figure 2 for detailed descriptions.

**Figure 6. Attention mechanism highlights patches important for the classification between IDH1-mutate and IDH1-wild patients in GBM.** For all IDH1-mutate patients in GBM, we selected top four patches with the highest attention weights for patient with top predicted probability of mutation then repeated the same process for wild-type patients. **(a)** We visualize them by combining WSI thumbnails, scatterplots of best clusters, and Attention weight heatmaps in figure on the left. The purple scatter points represent the patches in the optimal cluster, and the scatter points in the heat bar represent the attention weights of the representative patches selected by the model. The redder the color, the higher the weight. **(b)** The four different color

scatter plots in the middle represent the scatter points of the four clusters. (**c**) The four figures on the right are top four patches with the highest attention weights, which were annotated and analyzed by a pathologist.

**Figure 7**. **Attention mechanism highlights patches important for the classification between BAP1-mutate and BAP1-wild patients in KIRC.** Refer to the caption of Figure 6 for detailed descriptions.

**Table 1. Analysis result of AMIML and some existing methods for several candidate genes based on TCGA cohorts.**

**Figure 1.**

(a)

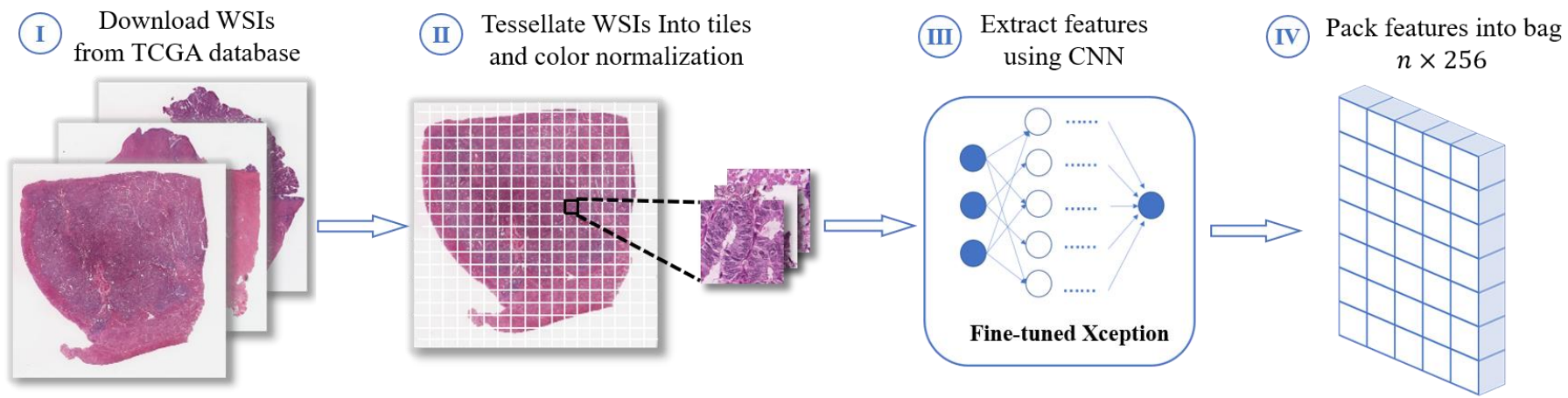

(b)

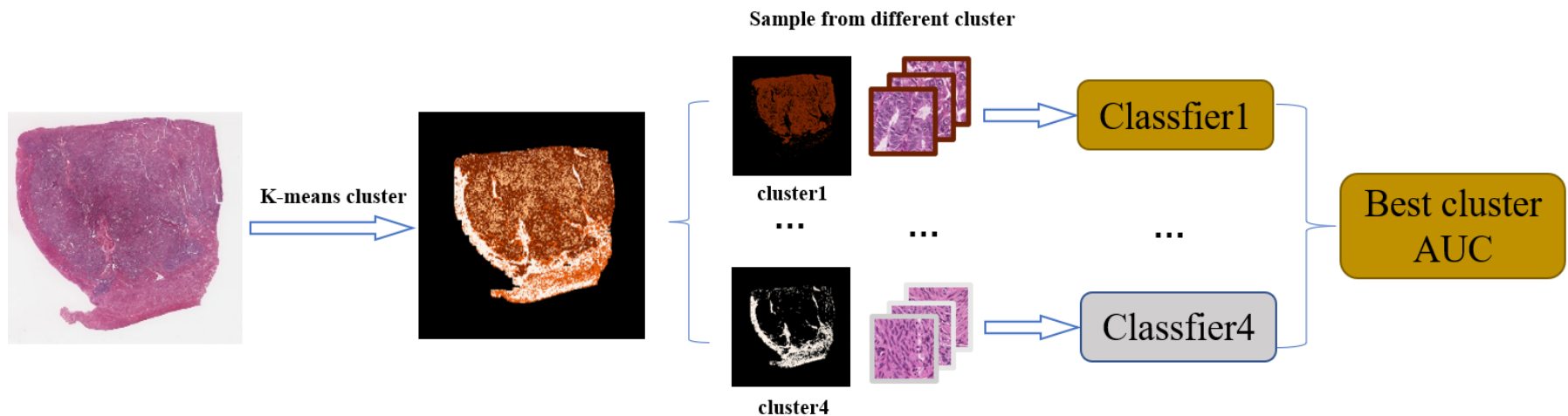

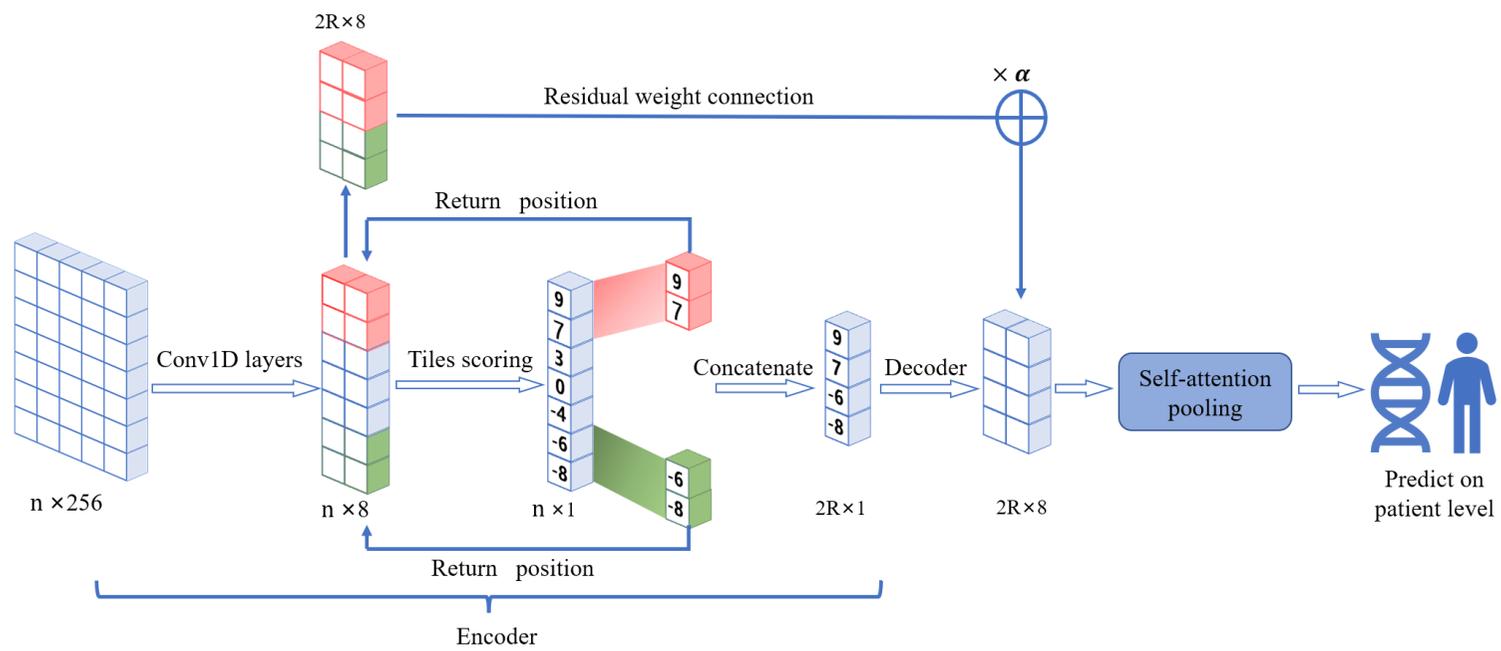

**Figure 2.**

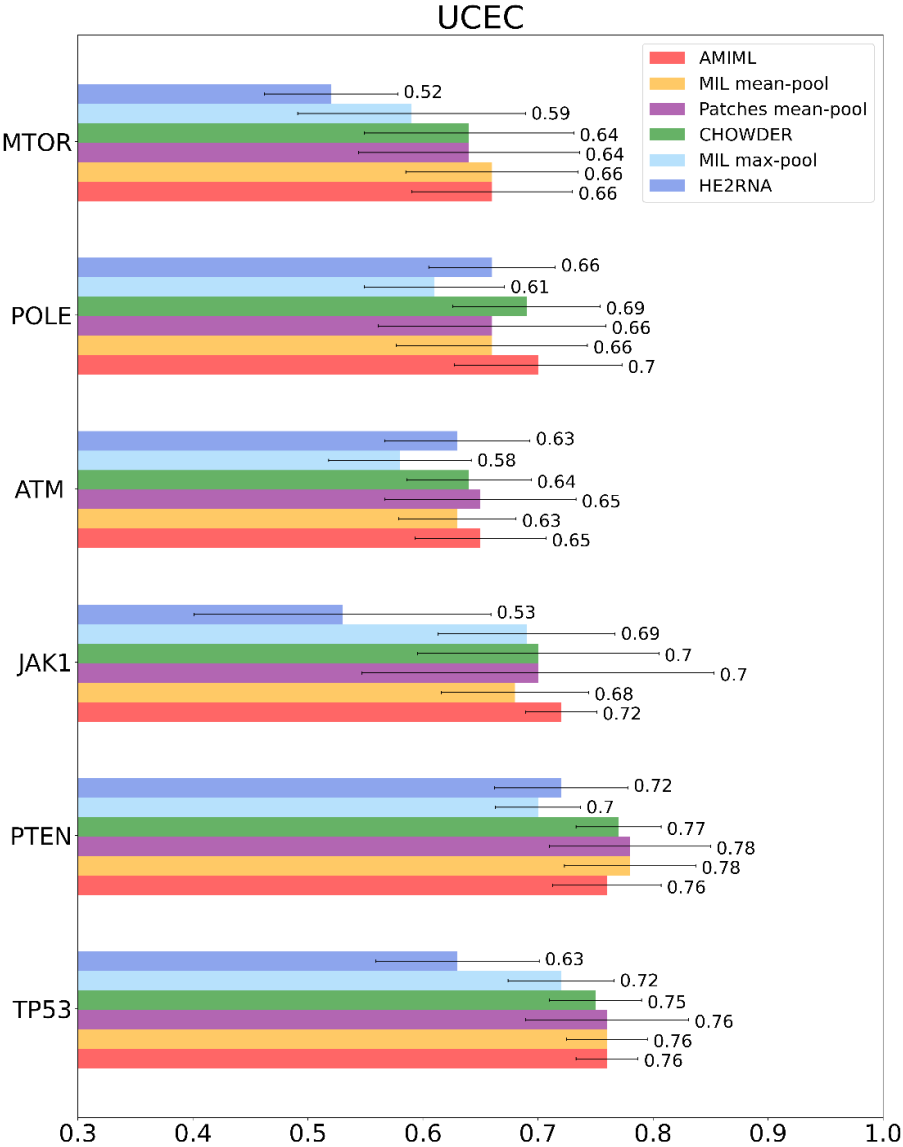

**Figure 3.**

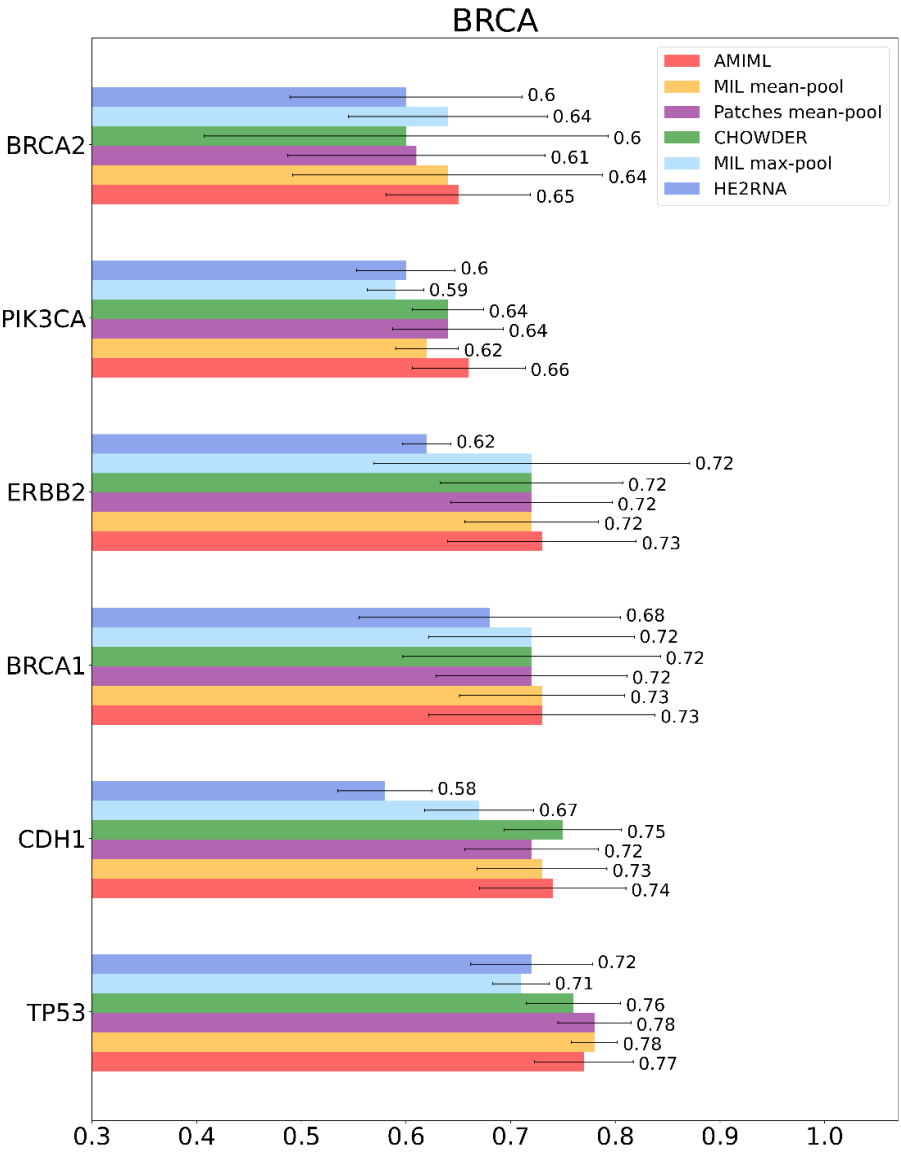

**Figure 4.**

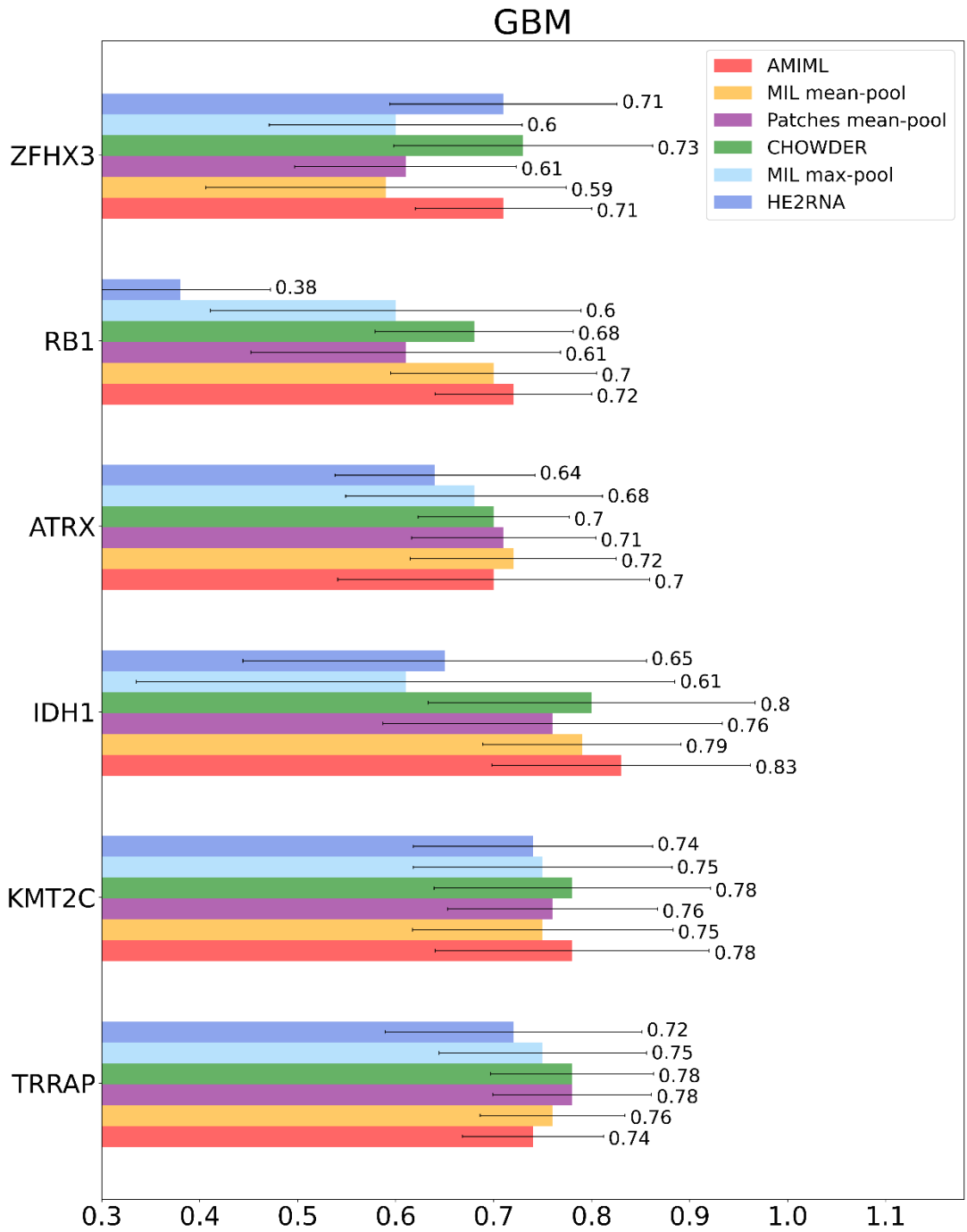

**Figure 5.**

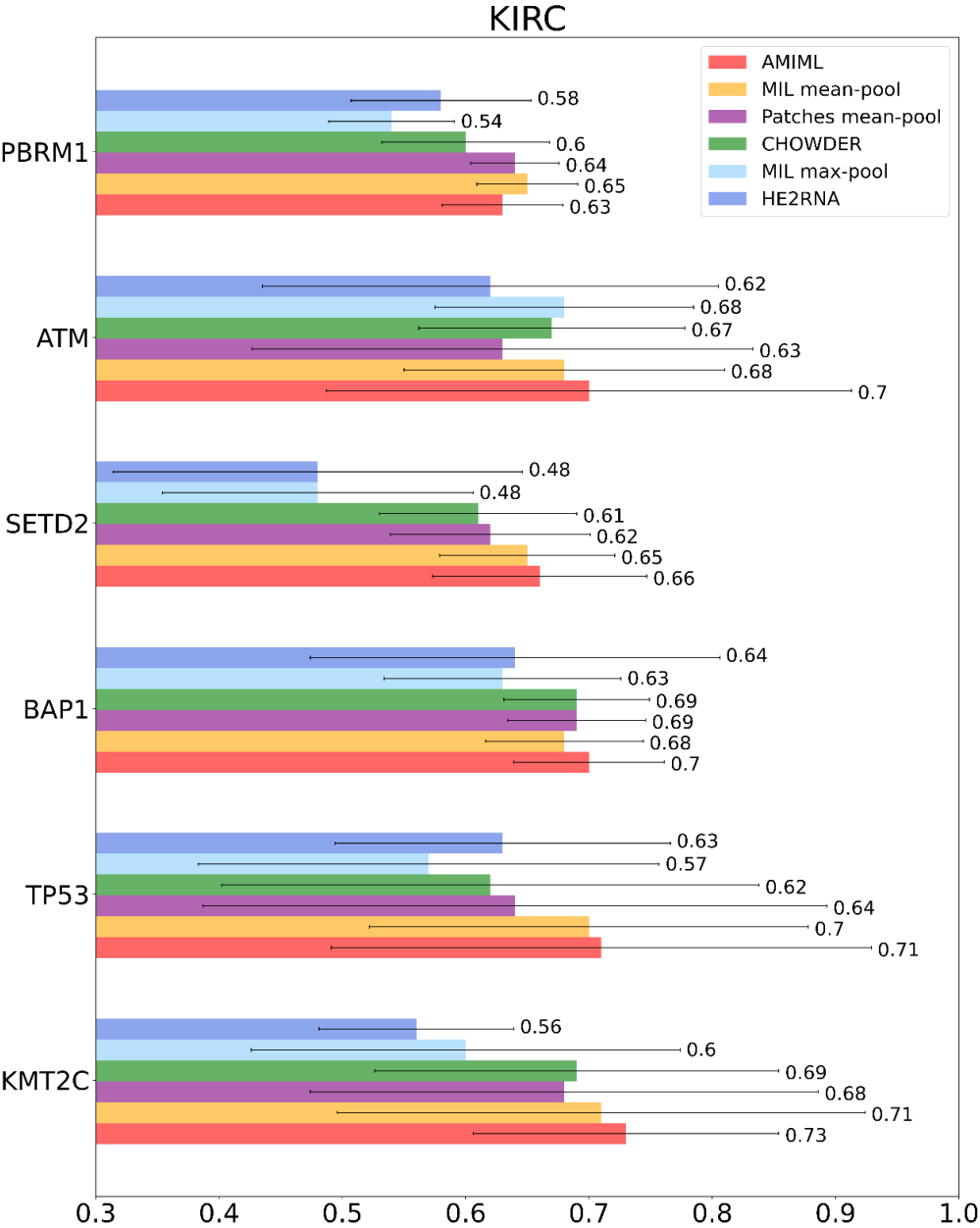

**Figure 6.**

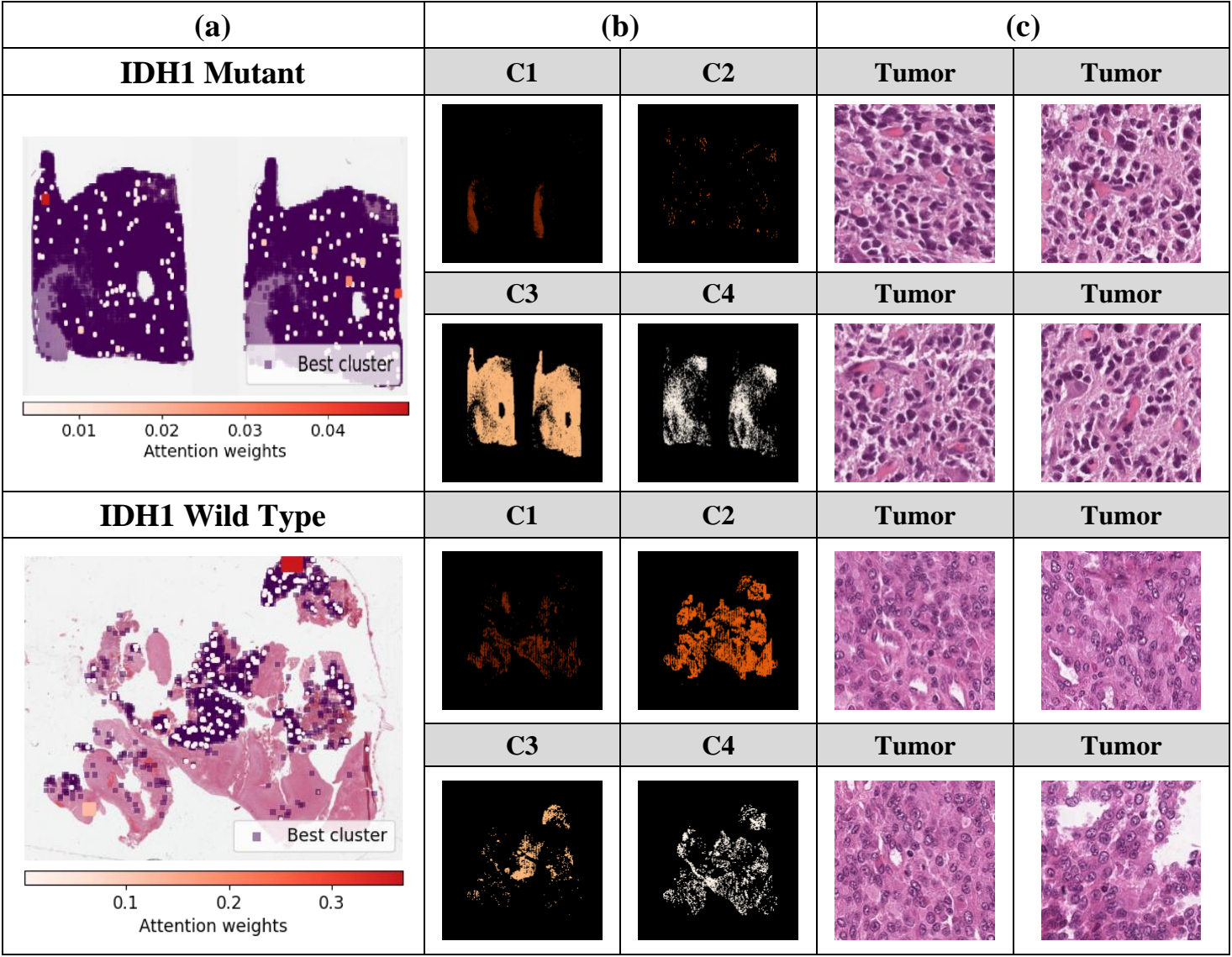

**Figure 7.**

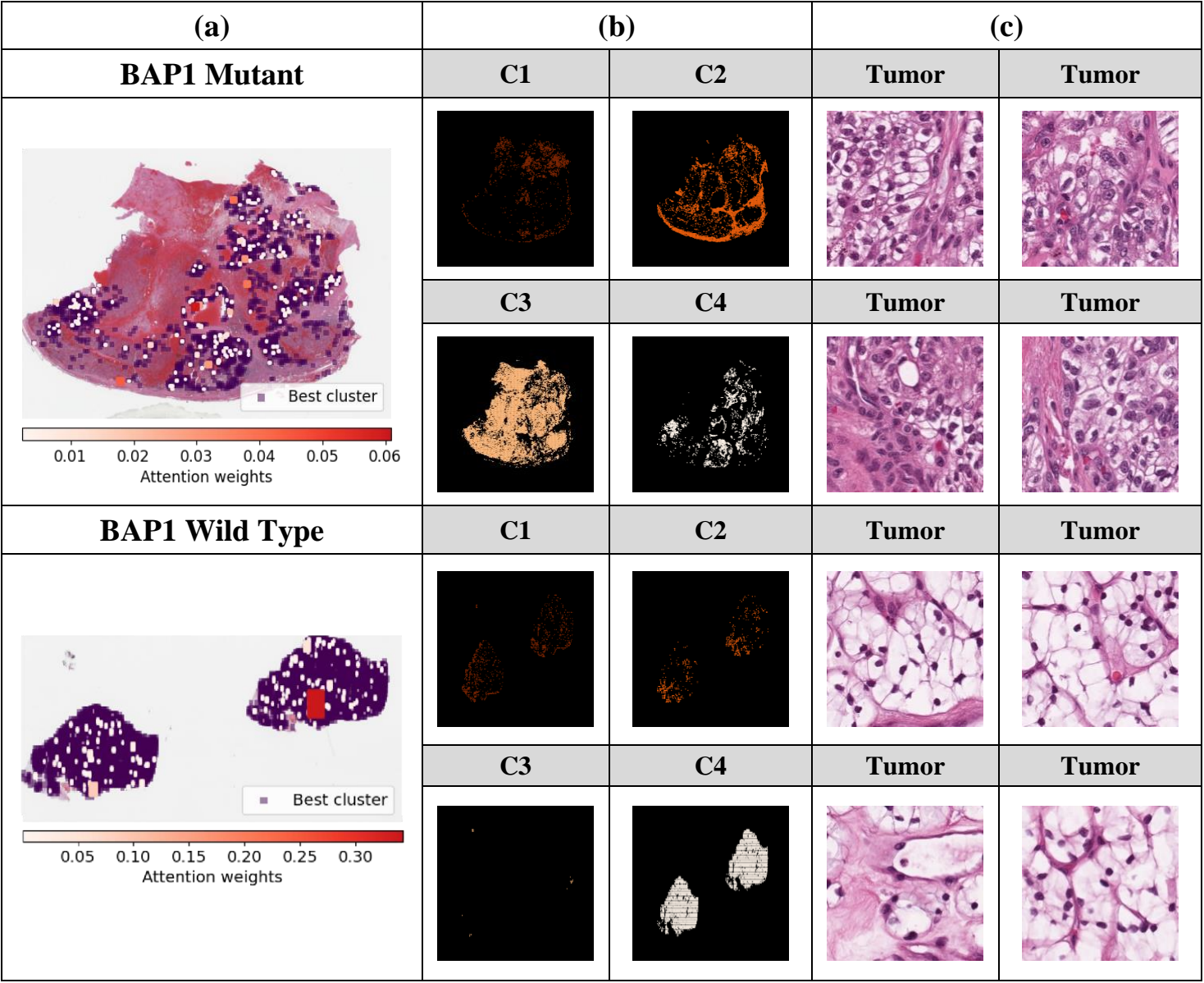

**Table 1**.

| Gene | AMIML | | | | Existing methods | | | | |
|---|---|---|---|---|---|---|---|---|---|
| | Total | Mutant | Wild Type | AUC | Total | Mutant | Wild Type | AUC | Reference |
| **UCEC** | | | | | | | | | |
| TP53 | 388 | 147 | 241 | 0.76 | 361+98 | | | 0.87 | Hong et al.[42] |
| PTEN | 388 | 250 | 138 | 0.76 | 361+98 | | | 0.77 | Hong et al. |
| JAK1 | 388 | 53 | 335 | **0.72** | 361+98 | | | 0.61 | Hong et al. |
| ATM | 388 | 71 | 307 | 0.65 | 361+98 | | | 0.65 | Hong et al. |
| POLE | 388 | 62 | 326 | **0.70** | 361+98 | | | 0.66 | Hong et al. |
| MTOR | 388 | 43 | 345 | **0.66** | 361+98 | | | 0.61 | Hong et al. |
| Note: Hong et al. used a combined dataset of TCGA and CPTAC, Sample size of TCGA+CPTAC=361+98 | | | | | | | | | |
| **BRCA** | | | | | | | | | |
| TP53 | 728 | 249 | 479 | 0.77 | 995 | 325 | 670 | 0.78 | Kather et al.[1] |
| | | | | | 985 | 338 | 647 | 0.75 | Noorbakhsh et al.[18] |
| CDH1 | 728 | 85 | 643 | 0.74 | 659 | 65 | 594 | 0.78 | Qu et al[30] |
| BRCA1 | 728 | 20 | 708 | **0.73** | 995 | 26 | 969 | 0.56 | Kather et al. |
| ERBB2 | 728 | 23 | 705 | **0.72** | 995 | 27 | 968 | 0.64 | Kather et al. |
| | | | | | 659 | 24 | 625 | 0.63 | Qu et al |
| PIK3CA | 728 | 258 | 470 | **0.66** | 995 | 322 | 673 | 0.63 | Kather et al. |
| | | | | | 659 | 203 | 456 | 0.58 | Qu et al |
| BRCA2 | 728 | 19 | 708 | **0.65** | 995 | 28 | 967 | 0.52 | Kather et al. |
| **GBM** | | | | | | | | | |
| TRRAP | 214 | 16 | 198 | **0.79** | Deregulation of this gene may play a role in several types of cancer including glioblastoma multiforme. | | | | |
| KMT2C | 214 | 15 | 199 | **0.78** | This gene is a member of the myeloid/lymphoid or mixed-lineage leukemia (MLL) family | | | | |
| IDH1[44] | 214 | 14 | 200 | 0.83 | LGG+GBM | 682 slides | 439 slides | 0.84 | Cui et al.[43] |
| ATRX[50] | 214 | 20 | 194 | **0.73** | | | | | |
| RB1[44] | 214 | 23 | 191 | **0.72** | The protein encoded by this gene is a negative regulator of the cell cycle and was the first tumor suppressor gene found. | | | | |
| ZFHX1 | 214 | 14 | 200 | **0.73** | This gene is reported to function as a tumor suppressor in several cancers | | | | |
| **KIRC** | | | | | | | | | |
| KMT2C | 317 | 13 | 304 | **0.73** | 481 | 14 | 467 | 0.63 | Kather et al.[1] |
| TP53 | 317 | 8 | 309 | **0.71** | 481 | 11 | 470 | 0.62 | Kather et al. |
| BAP1[46] | 317 | 32 | 285 | **0.70** | 481 | ... | ... | ... | |
| SETD2 | 317 | 39 | 278 | **0.66** | 481 | 47 | 434 | 0.62 | Kather et al. |
| ATM | 317 | 15 | 302 | **0.70** | 481 | 12 | 469 | 0.69 | Kather et al. |
| PBRM1 | 317 | 135 | 182 | **0.63** | 481 | 147 | 33 | 0.63 | Kather et al. |

# Supplementary

**(a)**

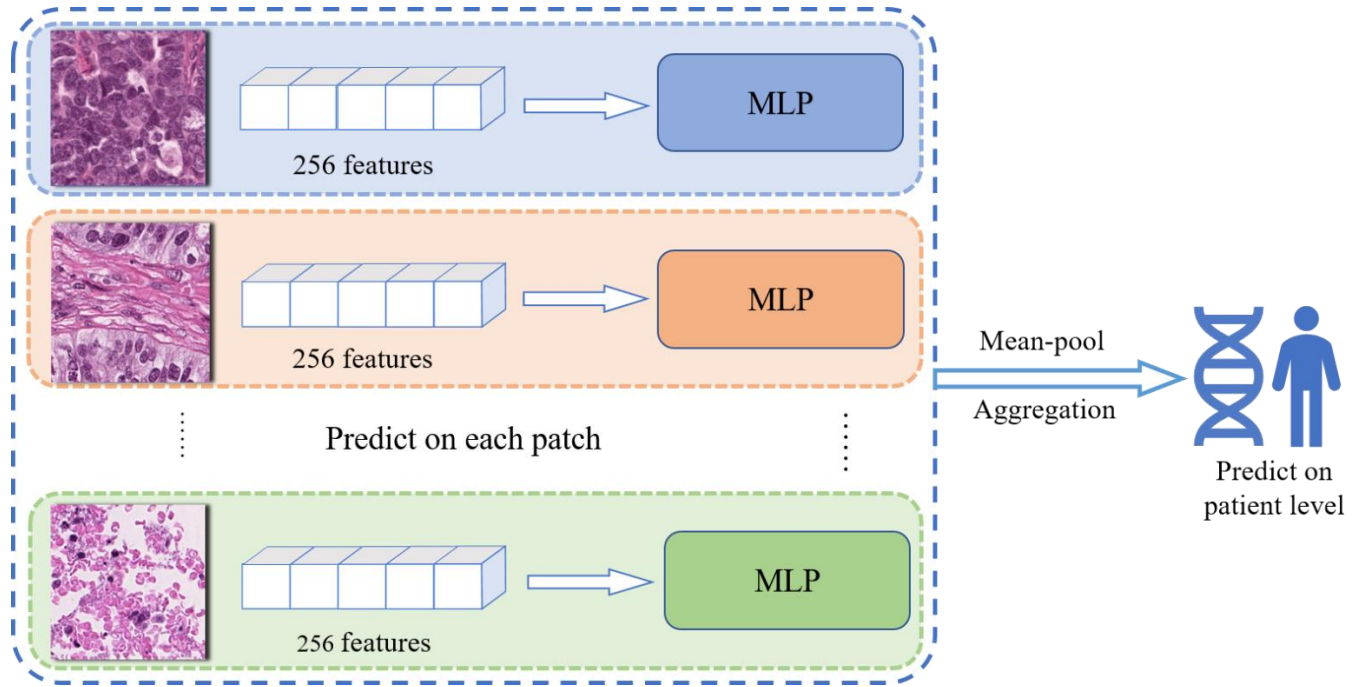

**(b)**

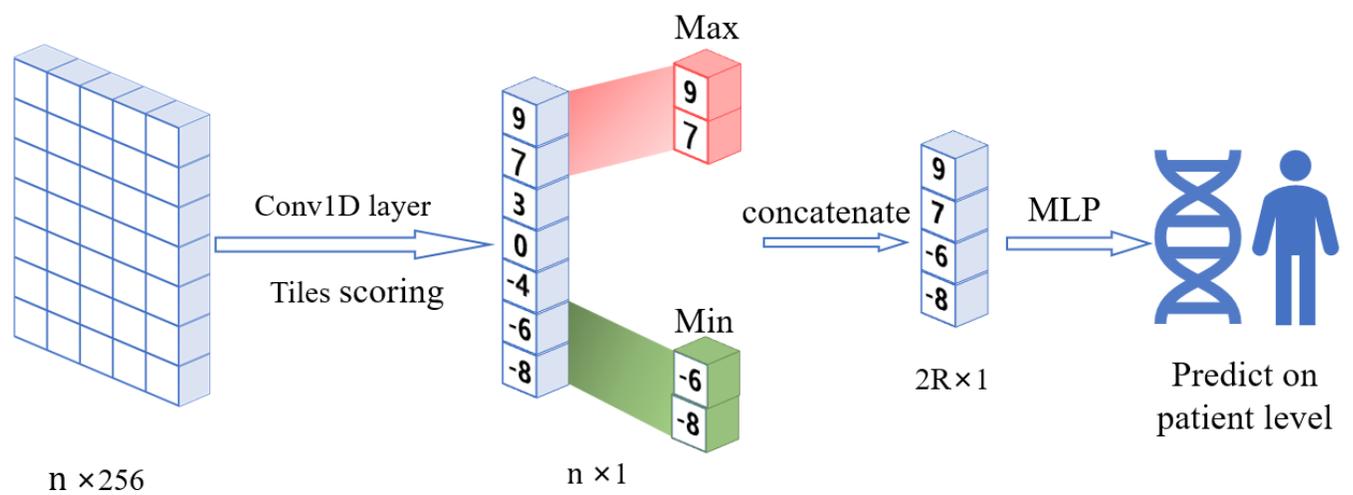

**(c)**

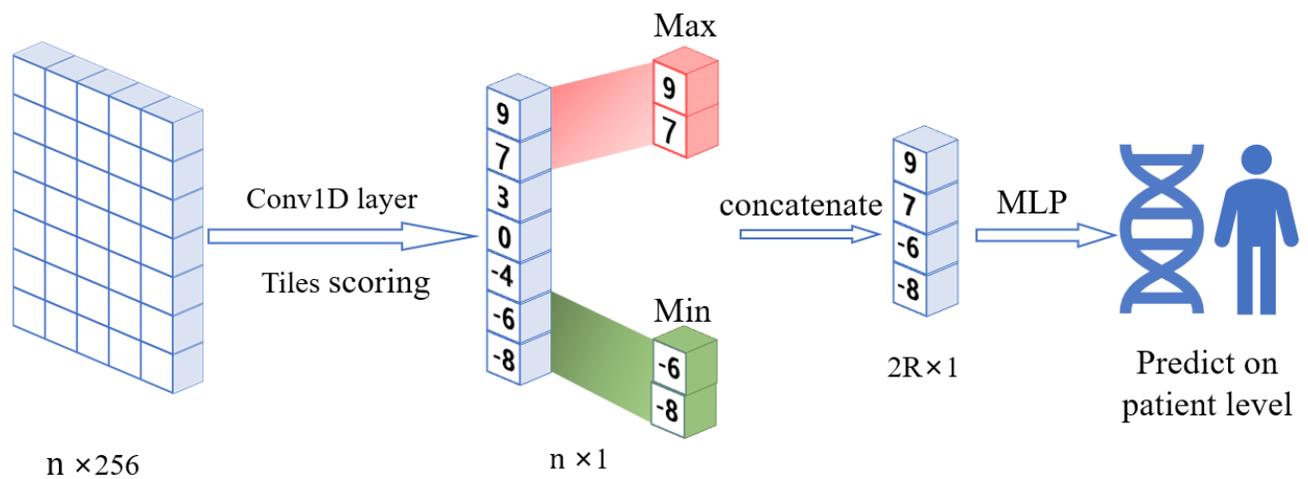

**Supplementary Figure 1 The architecture of baseline algorithms. (a)** The architecture of patches mean-pool aggregation method. In the first step, the features of each patch are input into a simple multilayer perceptron (MLP) to predict patch-level prediction; In the second step, the patient-level prediction is obtained by averaging the outputs of patches of this patient. **(b)** The architecture of the CHOWDER method in the case of R = 2. **(c)** The architecture of the MIL mean-pool method.

| Cluster1 | Cluster2 | Cluster3 | Cluster4 |
|---|---|---|---|
| 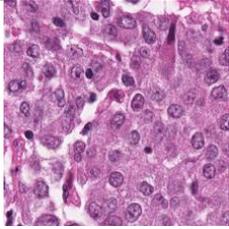 | 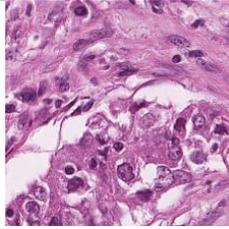 | 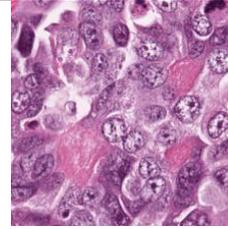 | 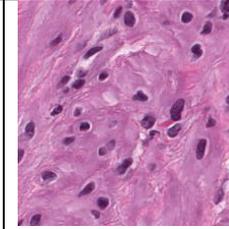 |
| Tumor | Tumor | Tumor | Stroma |
| 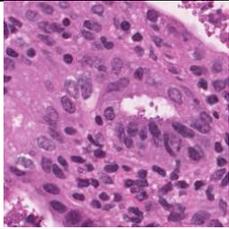 | 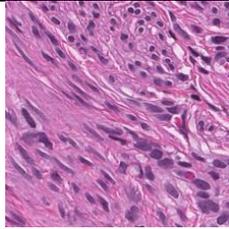 | 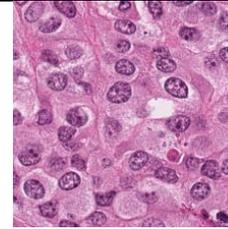 | 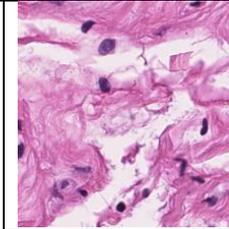 |
| Endometrial glands and stroma | Stroma | Tumor | Connective tissue |
| 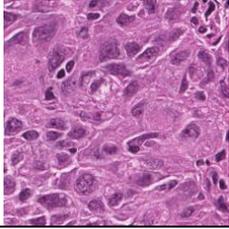 | 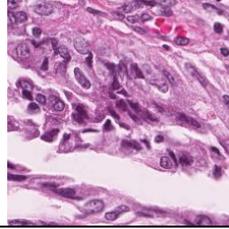 | 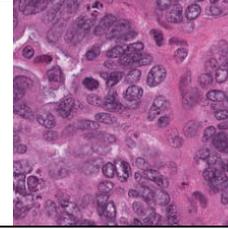 | 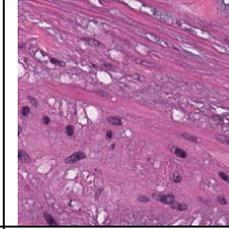 |
| Tumor | Tumor | Tumor | |
| 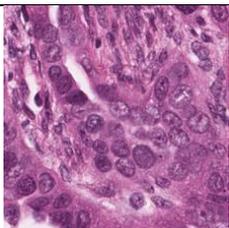 | 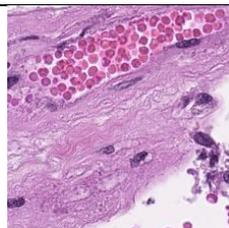 | 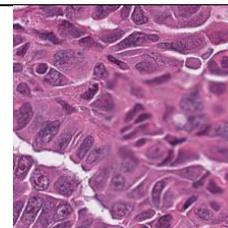 | 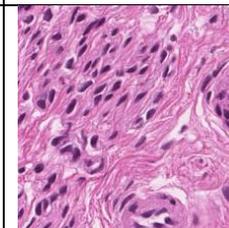 |
| Tumor | Blood vessel | Tumor | Stroma |

**Supplementary Figure 2. Representative Image patches for each cluster for UCEC.** In order to find the most representative patches in each cluster, we used the K-means method to cluster each cluster into 4 subclusters, and identified the patch closest to the center of each subcluster. Four representative patches were obtained for each cluster in this method. These patches were then annotated by a pathologist.

| Cluster1 | Cluster2 | Cluster3 | Cluster4 |
|---|---|---|---|
| 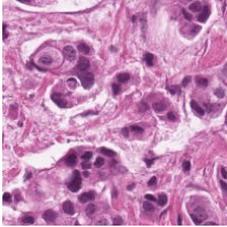 | 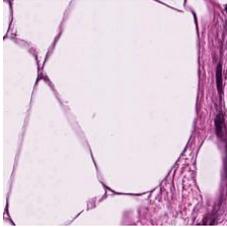 | 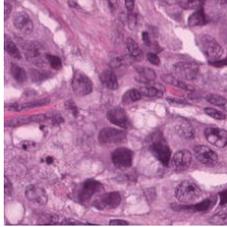 | 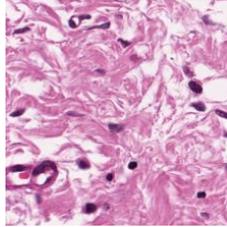 |
| Tumor | Adipose | Tumor | Loose connective tissue |
| 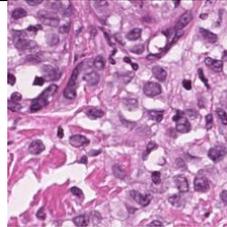 | 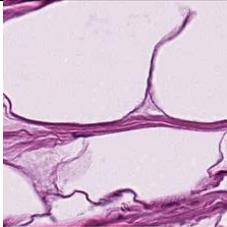 | 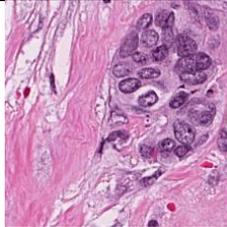 | 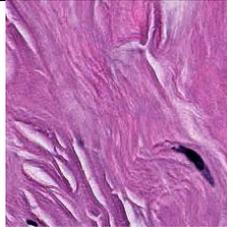 |
| Tumor | Adipose | Tumor | Fibrous tissue |
| 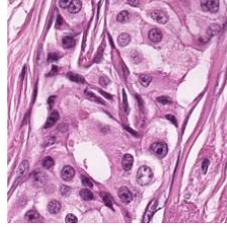 | 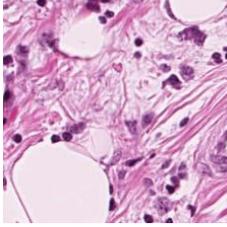 | 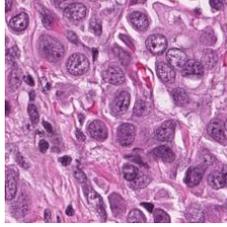 | 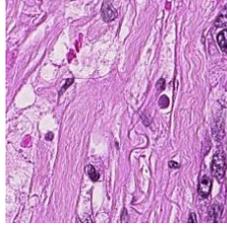 |
| Tumor | Loose connective tissue | Tumor | Stroma |
| 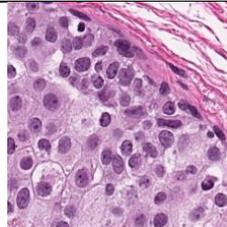 | 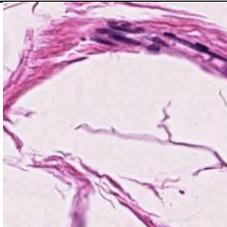 | 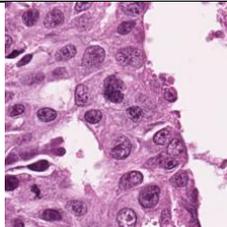 | 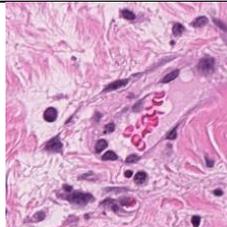 |
| Tumor | Adipose | Tumor | Blood vessel |

**Supplementary Figure 3. Representative Image patches for each cluster for BRCA.**

| Cluster1 | Cluster2 | Cluster3 | Cluster4 |
|---|---|---|---|
| 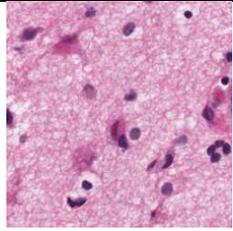 | 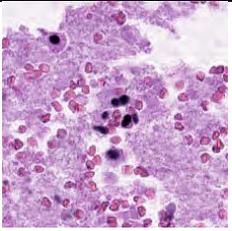 | 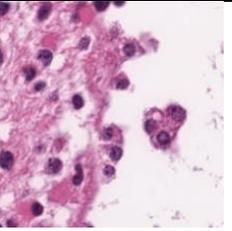 | 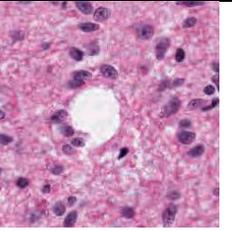 |
| Neurons and gliocytes | Neurons, gliocytes and red cells | Neurons and gliocytes | Neurons and gliocytes |
| 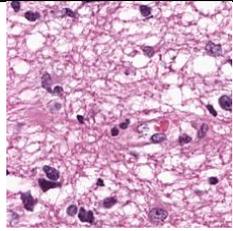 | 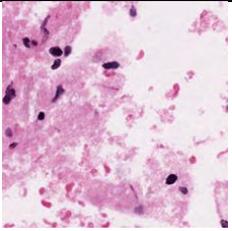 | 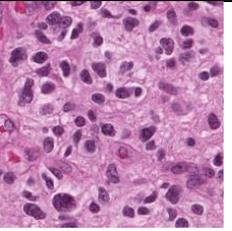 | 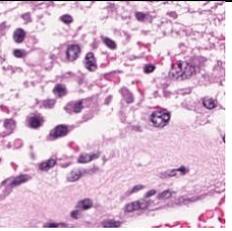 |
| Neurons and gliocytes | Neurons and gliocyte | Tumor | Tumors |
| 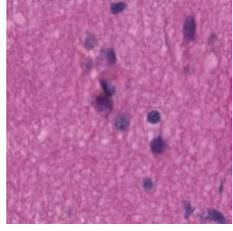 | 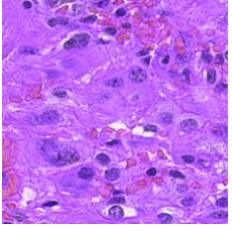 | 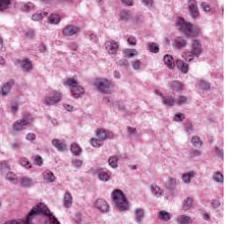 | 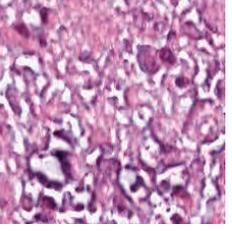 |
| Neurons and gliocytes | Tumor | Tumor | Tumors |
| 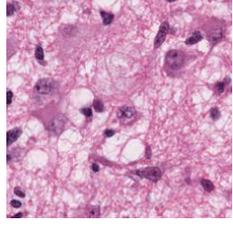 | 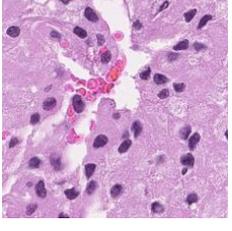 | 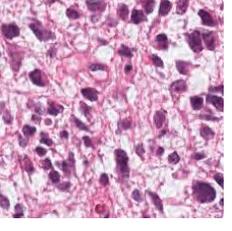 | 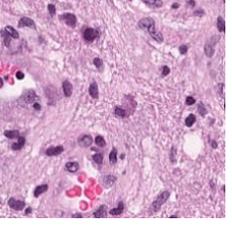 |
| Tumor | Neurons and gliocytes | Tumor | Neurons and gliocytes |

**Supplementary Figure 4. Representative Image patches for each cluster for GBM.**

| Cluster1 | Cluster2 | Cluster3 | Cluster4 |
|---|---|---|---|
| 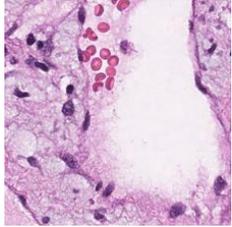 | 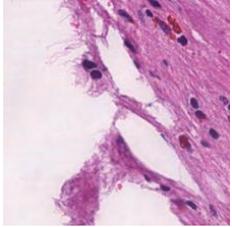 | 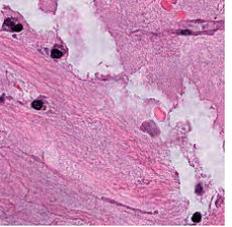 | 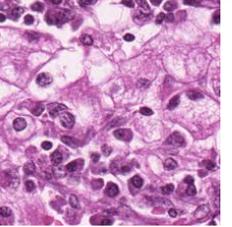 |
| Tumor | Cyst wall | Cyst wall | Tumor |
| 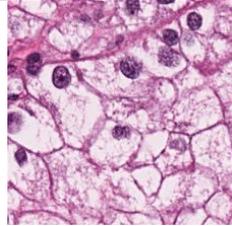 | 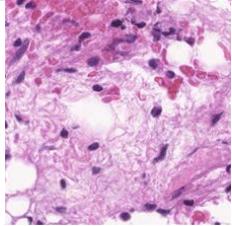 | 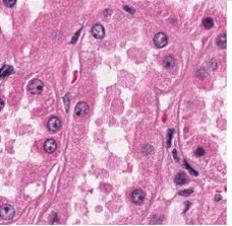 | 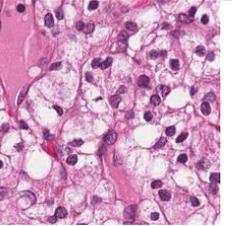 |
| Tumor | Loose connective tissue | Renal tubule | Tumor |
| 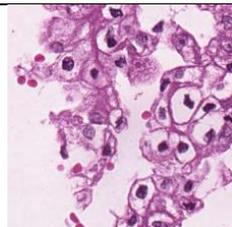 | 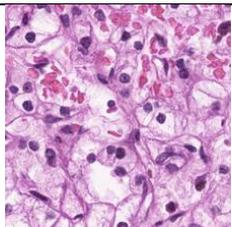 | 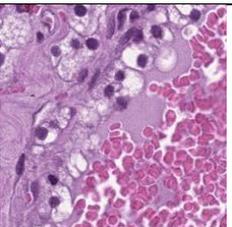 | 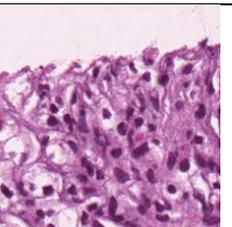 |
| Tumor | Tumor | Tumor and red cells | Tumor |
| 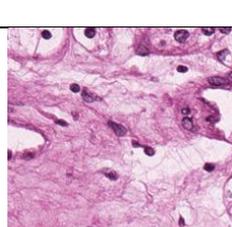 | 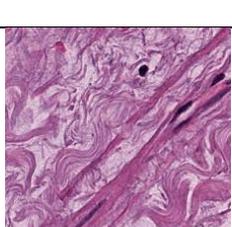 | 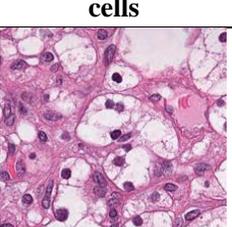 | 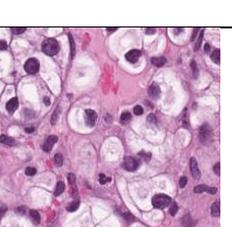 |
| Loose connective tissue | Loose connective tissue | Tumor | Tumor |

**Supplementary Figure 5. Representative Image patches for each cluster for KIRC.**

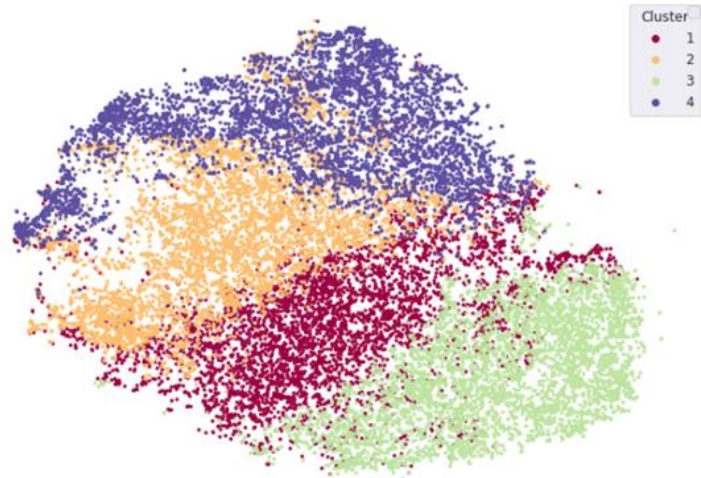
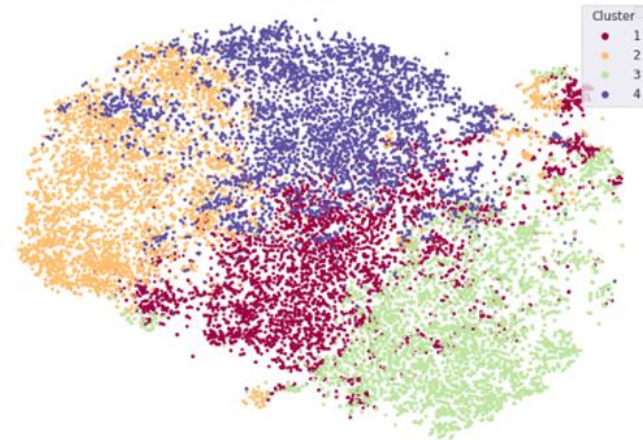
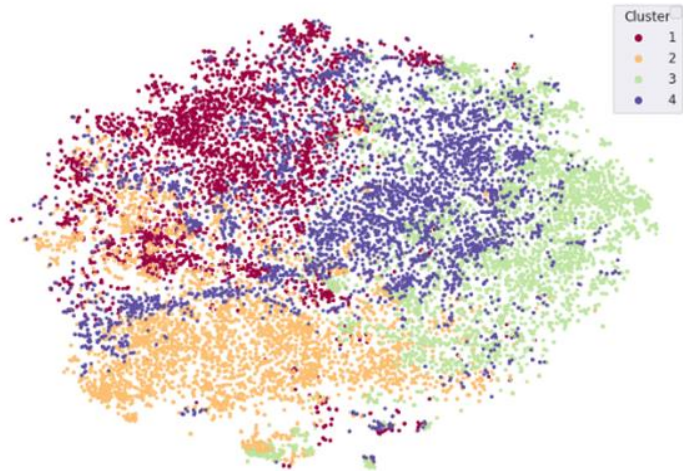
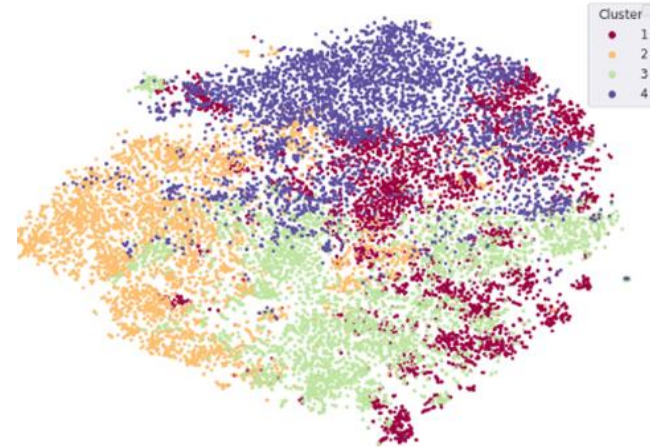

**Supplementary Figure 6. The visualization of clustering results by t-SNE.** For each cancer, 5,000 patches were randomly selected from each of the four clusters and displayed using t-distributed stochastic neighbor embedding (t-SNE) dimensionality reduction representation.